# NUMERICAL FORECAST OF THE MELTING AND THERMAL HISTORIES OF PARTICLES INJECTED IN A PLASMA JET

Jorge Romero Rojas[1], Marcela A. Cruchaga[2], Diego J. Celentano[3], Mohammed El Ganaoui[4] ,Bernard Pateyron[4]

1 Departamento de Ingeniera Metalurgica. Universidad de Santiago de Chile. Santiago, Chile.
2 Departamento de Ingeniera Mecanica. Universidad de Santiago de Chile. Av. Bdo. O'Higgins 3363. Santiago, Chile. E-mail: marcela.cruchag@usach.cl
3 Departamento de Ingeniera Mecanica - Metalurgica. Pontificia Universidad Catolica de Chile. Vicuaa Mackenna 4860. Santiago, Chile.
4 Faculté des Sciences. Universita de Limoges, 123 Albert Thomas - 87000. Limoges, France.

## ABSTRACT

*This work presents the numerical simulation of the melting process of a particle injected in a plasma jet. The plasma process is nowadays applied to produce thin coatings on metal mechanical components with the aim of improving the surface resistance to different phenomena such as corrosion, temperature or wear. In this work we studied the heat transfer including phase-change of a bi-layer particle composed of a metallic iron core coated with ceramic alumina, inside a plasma jet. The model accounted for the environmental conditions along the particle path. The numerical simulation of this problem was performed via a temperature-based phase-change finite element formulation. The results obtained with this methodology satisfactorily described the melting process of the particle. Particularly, the results of the present work illustrate the phase change evolution in a bi-layer particle during its motion in the plasma jet. Moreover, the numerical trends agreed with those previously reported in the literature and computed with a finite volume enthalpy based formulation.*

*Keywords: Phase-change, heat transfer, bi-layer particle, plasma.*

## INTRODUCTION

Coating is a very important technique to improve surface properties, especially for cases with extreme working conditions such as corrosion, high temperature and wear. A protective layer of coating is typically applied to provide a barrier which prolongs component durability and the desired properties of hardware components in electronic and other engineering devices. Among the different coating systems, the thermal barrier coatings (TBCs) are commonly used to protect hardware operating in high temperature environments, such as combustor liners and gas turbine blades, from excessively high heat fluxes and temperatures. The most commonly used thermal spray TBC materials for turbine engines are oxide ceramics, among them yttria-stabilized zirconia (YSZ) and alumina alloys. However, other materials can be used for thermal barriers such as metal alloys.

Thermal barrier coatings are conventionally applied by introducing a powder of the coating material into a plasma jet in which powder particles are melted and accelerated towards the surface to be coated. While this technology has matured over the past several decades, the recent developments have focused on attaining nanometer size features in the coating microstructure for superior coating properties in terms of better service performance and spallation resistance [1].

Due to the inherent difficulty to measure the temperature in a moving particle, numerical simulations become relevant to describe its temperature and melted fraction evolutions for a better understanding and control of the process [2].

The simulation of particles behavior during its paths inside the plasma jet is a relative new topic. Many models have been developed in order to simulate the conditions of plasma gases using fluid mechanics and heat transfer analyses. These approaches consider type and flow of gases, type and material of nozzles, power and other parameters related to plasma coating processing [3,4,5]. Others works have based their studies on simulate and estimate mechanical properties and thermal evolution of coating according to the plasma process parameters [6,7,8]. Latest works have studied thermal evolution for ceramic, metallic or ceramic with solvent liquids in-flight particles by means of models or softwares which provide plasma gases properties [9, 10,11,12,13,14,15,16], together with coating micro-structural evolution [17,18]. All these models consider isothermical particles, neglecting heat transfer inside them. More recent works [19,20] analyze the heat transfer with phase change in particles composed of one material (porous zirconia in [19] and in ceria [20]) using a mixture of argon hydrogen as plasma gases. In the present study we analyze the thermal evolution including phase change effects of bi-layer particles, composed by metallic iron core surrounded by a covering layer of ceramic alumina in a plasma gas similar to those used by [19,20].

The heat transfer problem inside the particle is analyzed using a temperature-based phase-change finite element formulation presented and validated in many applications in [21, 22,23,24,25]. The gas conditions along the particle path needed as input for this simulation are previously computed with an *ad-hoc* software [26].

## THEORETICAL ANALYSIS

### Thermal barrier coating

Modern TBC's are required to not only limit heat transfer through the coating but to also protect engine components from oxidation and hot corrosion. No single coating composition appears able to satisfy these multifunctional requirements. As a result, a "coating system" has evolved. Research in the last 20 years has led to a preferred coating system consisting of three separate layers [27] to achieve long term effectiveness in the high temperature, oxidative and corrosive use environment for which they are intended to function (**Figure 1**).

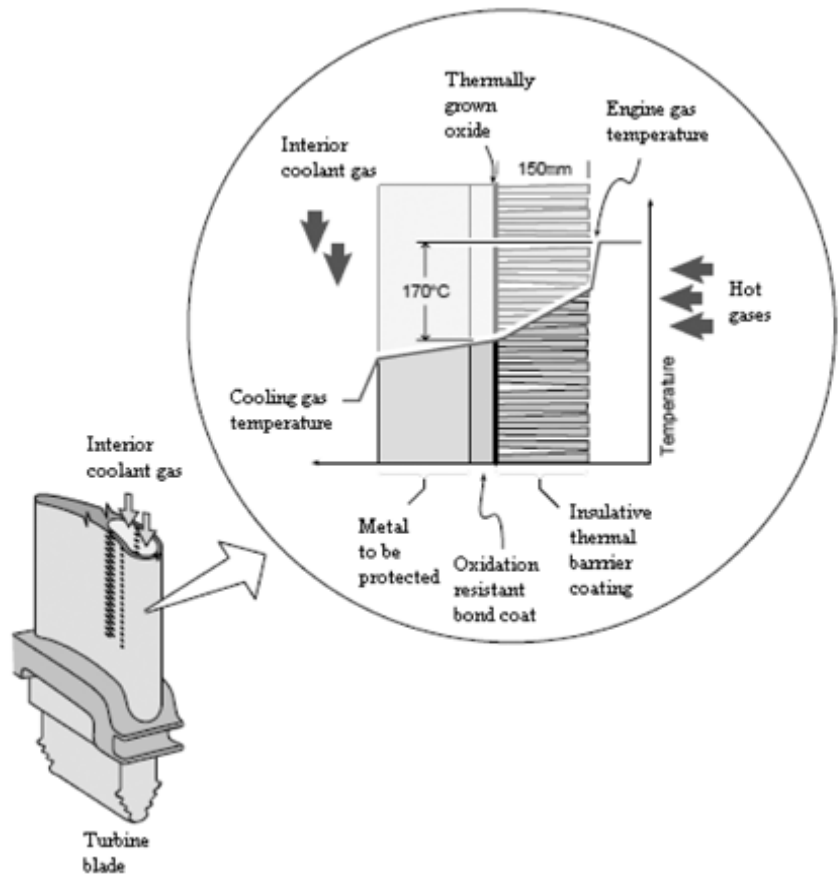

**Figure 1** - A schematic illustration of a modern thermal barrier coating system consisting of a thermally insulating thermal barrier coating, a thermally grown oxide (TGO) and an aluminum rich bond coat. The temperature gradient during engine operation is overlaid.

First, a thermally protective TBC layer with a low thermal conductivity is required to maximize the thermal drop across the thickness of the coating. This coating is likely to have a thermal expansion coefficient that differs from the component to which it is applied. This layer should therefore have a high in-plane compliance to accommodate the thermal expansion mismatch between the TBC and the underlying metal substrate component. In addition, it must be able to retain this property and its low thermal conductivity during prolonged environmental exposure. A porous, columnar, 100-200 μm thick, yttria stabilized zirconia (YSZ) layer is currently preferred for this function [28]. This layer may be applied using either air plasma spray (APS) or electron beam physical vapor deposition (EB-PVD). Second, an oxidation and hot corrosion resistant layer is required to protect the underlaying turbine blade from environmental degradation. This layer is required to remain relatively stress free and stable during long term exposure and remain adherent to the substrate to avoid premature failure of the TBC system. It is important that it also provide an adherent surface for the TBC top coat. Normally, the thin (< 1 μm), protective aluminum rich oxide which is thermally grown upon the bondcoat is utilized for this purpose [29]. In addition, these layers are desired to be thin and low density to limit the centrifugal load on rotating engine components and have good thermal and mechanical compatibility.

**Production of bi-layer particles**

Bi-layer or "cermet" particles can be made by means of different techniques, the most common and simplest is "mechanofusion". This is a dry process where a powder mixture is subjected to various mechanical forces as it passes through a variable gap (few millimeters or micrometers) in a rotating device (Figure 2-a). As a result, small guest particles can be coated and bonded onto the surfaces of larger host particles (in this work alumina and iron, respectively) without using binder of any kind (Figure 2-b). The degree of coverage depends on: (i) the particle size ratio of the host to guest powders (typically a host:guest ratio of >10:1 is required); (ii) the proportion of the guest material; (3) the relative adhesive properties of the host and guest particles and (iv) the energy input from the machine. An excellent review of mechanofusion and similar dry coating techniques is presented in [30].

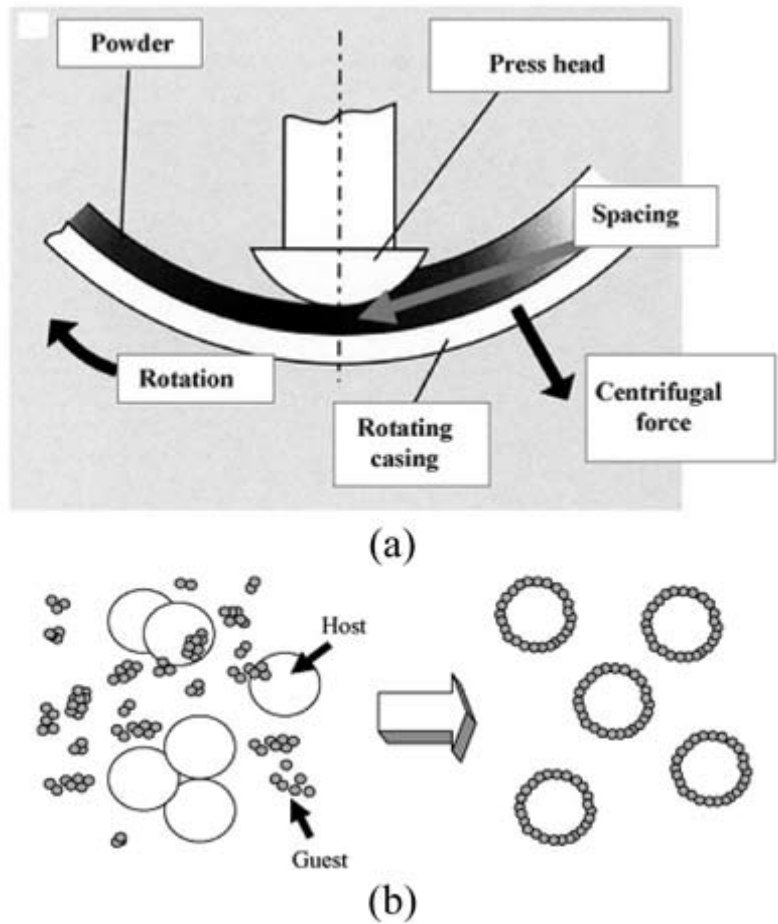

Figure 2 - Principle of the mechanofusion process.

### Plasma spraying

Plasma spraying is a process where metallic and non-metallic (ceramics) powders are deposited in a molten or semi-molten state on a prepared substrate. Typical thermal plasma heat source uses a direct current (DC) arc with temperatures over 8000 K at atmospheric pressure, allowing the melting of any material.

Powdered materials are injected inside the plasma jet where particles are accelerated and melted, or partially melted before they flatten and solidify onto the substrate, forming the splats. Thus, the coating thickens by the layering of splats (Figure 3 and Figure 4).

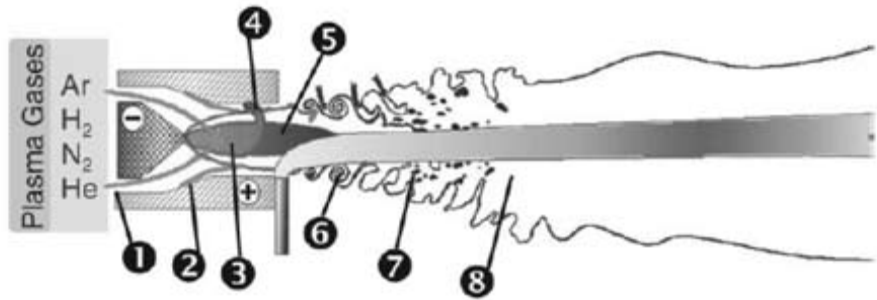

**Figure 3 -** Conventional DC arc spray torch. (1) nozzle, (2) plasma generation by gas injection, (3) cold boundary layer at the anode wall, (4) arc column, (5) connecting arc column, (6) plasma jet exiting the nozzle, (7) large scale eddies, (8) surrounding atmosphere bubbles entrained by the engulfment process.

This type of plasma torch has gas velocities between 600 m/s and 2300 m/s with a radial particles injection. The cathode is made of thoriated (2 wt %) tungsten and the anode-nozzle of high purity oxygen free copper (OFHP), often with an insert of sintered tungsten. Table 1 summarizes the main characteristics of DC plasma torches. According to the particle velocity, its residence times in the plasma jet core and its plume are between 0.1 s and a few milliseconds. Besides, particles should be injected at least with a momentum similar to that of the plasma jet. This is achieved by using a carrier gas or liquid with an injector whose internal diameter is between 1.2 mm and 2 mm.

The melting of a particle for a given size depends on two parameters: the residence time linked to its velocity inside the plasma, and the plasma gas composition. The latter mainly controls the heat transfer to particles, which is the lowest for pure Ar and the highest for ternary mixtures such as Ar-He-$H_2$, with intermediate values for Ar-He and Ar-$H_2$ mixtures. This device develops power levels between 35 kW and 50 kW, with nozzle diameters between 6 mm and 8 mm. For this type of particle injection, the last ones have a size smaller than 40 µm when the plasma gas is pure argon, and will require a secondary gas ($H_2$, He) to melt ceramic particles of the same size. When the particle size decreases, the gas velocity has to be drastically increased. For example, particles with a diameter within the range 5 µm -10 µm need to be injected mixed with a liquid (usually ethylene). In this case, the input flow rate drastically disturbs the plasma jet.

**Table 1 -** The main characteristics of DC arc spray torches.

| Type (anode material) | Plasma gas | Flow rate (slm) | Maximum arc current (A) | Maximum voltage (V) | Maximum power (kW) | Maximum powder flow rate (kg/h) |
|---|---|---|---|---|---|---|
| Conventional (OFHP copper or sintered W) | Ar<br>Ar- He<br>Ar-$H_2$<br>Ar-He-H2<br><br>N2-$H_2$ | 40- 100 | 1000<br>900<br>700<br>700<br>500 | 30<br>50<br>80<br>90<br>80 | 30<br>45<br>55<br>60<br>40 | 6-8 |
| High power (OFHP copper) | $N_2$ - $H_2$ | 200- 400 | 500 | 500 | 250 | 12-18 |
| Triplex (segmented OFHP copper) | Ar - He<br><br>Ar | 30-60 | 300 | 80-90 | 20- 55 | 6-10 |

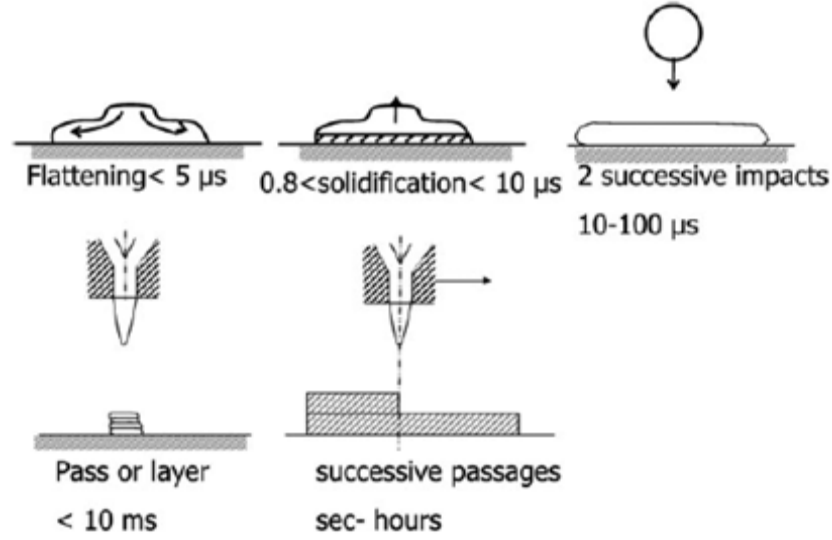

**Figure 4** - Intervals of time in coating formation.

## ANALYSIS OF THE PROBLEM

### Mathematical model

The heat transfer is governed by the well known energy equation:

$$\rho\left(c + L\frac{\partial f_{pc}}{\partial T}\right)\dot{T} = \nabla \cdot (k\nabla T) \qquad \text{in } \Omega \times Y \tag{1}$$

where $\Omega$ is an open-bounded domain with smooth boundary $\Gamma$, $Y$ is the time interval of interest ($t \in Y$), $\nabla$ is the gradient operator, $\rho$ is the material density, $c$ is the specific heat capacity, $L$ is the specific latent heat, $k$ is the conductivity coefficient and $f_{pc}$ is the phase-change function defined for a pure material as:

$$f_{pc} = H(T - T_m) \tag{2}$$

$H$ is the Heaviside function and $T_m$ is the melting temperature. In the context of the finite element method, the integral form of equation (1) is computed applying a temperature-based formulation to properly describe the latent heat release avoiding the use of any regularization in the phase-change function definition (i.e., the unit step function provided by the expression (2) is not smoothed via polynomial approximations; see [21-23] and references therein for further details).

### Numerical simulation

In this work we study the heat transfer including phase-change of a bi-layer particle composed of a metallic iron core surrounded by a covering layer of ceramic alumina (properties of particles are described in **Table 2**). Two different bi-layer spherical particles are utilized for the simulations considering constant and variable heat transfer coefficient to the environment (plasma gas). Thus, the four cases studied in this work are summarized in Table 3.

The particle path, the gas temperature and the convective heat transfer coefficient $h_\infty$ are obtained with the software Jets&Poudres [31] considering the weight of the particle, the injection angle and the plasma gas composition. The trajectory of the particle is described in Figure 5 together with the computed gas temperature field. The thermal evolution of the gas inside the plasma jet obtained using the software Jets&Poudres [32] is plotted in Figure 6. It can be seen that there is a uniform temperature between 0.3 and 0.4 ms, where the heat transfer coefficient is almost constant. This is the basic assumption for the analysis of our model; the two boundary temperatures being analyzed: around 10000 and over 11000 ÂºK.

The spraying particles processes takes place in short times, 1 ms as maximum, for this reason the convection coefficient $h_\infty$ is too high with the objective of promoting the complete melting of the particles before they splash against the substrate. Such coefficient depends on the temperature changes, Reynolds (*Re*) and Prandlt (*Pr*) numbers and other properties of the plasma gas varying with the particle path. The convection coefficient is given by:

$$h_\infty = \frac{Nu \cdot k_\infty}{d_p} \qquad (3)$$

with

$$Nu = \left[ 2 + 0.6 \, Re^{0.5} \, Pr^{0.33} \right] \cdot \left[ \frac{\rho_\infty \mu_\infty}{\rho_{ps} \mu_{ps}} \right]^{0.6} \qquad (4)$$

where $d_p$ is the particle diameter, *Nu* the Nusselt number and $\mu$ the viscosity. The subscript "∞" denotes bulk plasma gas properties and "*ps*" plasma gas on the particle surface. These properties are summarized in Table 4. From Figure 5, it can be seen that the particle path (yellow line) can be considered, according to [2], as an optimal particle trajectory since it is not deviated from the central axis for more than 4°. The computed plasma gas temperature and coefficient $h_\infty$ are assumed known data for the phase-change heat transfer model briefly presented in the former section. The finite element mesh utilized in the simulations 1, 2 and 3 is composed of 720 axisymmetric elements (Figure 7a), with 600 for the iron core and 120 for the ceramic layer (green and blue zones, respectively). For simulation 4, the mesh is composed of 600 elements for the iron core and 240 for the ceramic layer (Figure 7b). Besides, for both meshes 40 surface elements aimed at describing the particle-environment thermal interaction, were used.

Table 2 - Thermodynamic properties of the particles used for this work.

| Property | Solid Fe | Liquid Fe | Solid $Al_2O_3$ | Liquid $Al_2O_3$ | Solid bi-layer | Liquid bi-layer |
|---|---|---|---|---|---|---|
| **Density [kg/m$^3$]** | 7874 | 7874 | 3990 | 3990 | 7114 | 7114 |
| **Conductivity [W/m K]** | 30 | 55 | 5 | 6.30 | 28.25 | 51.59 |
| **Heat Capacity [J/kg K]** | 400 | 1000 | 1200 | 1888 | 488 | 1097 |
| **Latent Heat [J/kg]** | | 247211 | | 1093366 | | 340034 |
| **Melting Point [K]** | | 1810 | | 2326 | | 2000 |
| **Boiling Latent Heat [J/kg K]** | | $6.2*10^6$ | | $11.3*10^6$ | | $6.8*10^6$ |
| **Boiling Point [K]** | | 3135 | | 3800 | | 3500 |

Figure 8 presents the initial temperature conditions of the particle for cases 1 and 2. This temperature distribution is obtained considering that the particle has reached a thermal equilibrium in the plasma jet, remained 0.3 ms according to the results obtained in [1,31,32] for a thermal contact resistance "RTC" equals to $10^{-8}$ [m$^2$·K/W], which is the minimal value given for the author in this case. Characteristics and properties of plasma gases and particles were the same in both cases. Thus, thermal evolutions given by [1,33,34] and our model had the same initial conditions to evaluate and compare the phase change during its flight in the plasma gases. For cases 3 and 4 the initial conditions were the environmental ones (see Table 3). The non-uniform surface distribution of the heat transfer coefficient used in case 2 is given in Figure 9.

**Table 3 - Cases analyzed in the present work.**

| Properties | Case 1 | Case 2 | Case 3 | Case 4 |
|---|---|---|---|---|
| **Particle diameter [µm]** | 60 | 60 | 50 | 50 |
| **Plasma gas temperature [K]** | 11357 | 11357 | 10027 | 10027 |
| **Initial particle temperature [K]** | Given in Figure 6 | Given in Figure 6 | 300 | 300 |
| **$h_\infty$ [W/m$^2$ K]** | 66000 uniform | non-uniform given in Figure 7 | 55000 uniform | 55000 uniform |
| **$k_\infty$ [W/mÂ·K]** | 2.54 | 2.54 | 1.75 | 1.75 |
| **Time step [s]** | $1*10^{-6}$ | $1*10^{-6}$ | $1*10^{-7}$ | $1*10^{-7}$ |
| **$R_{Fe}/R_{particle}$** | 0.93 | 0.93 | 0.92 | 0.72 |

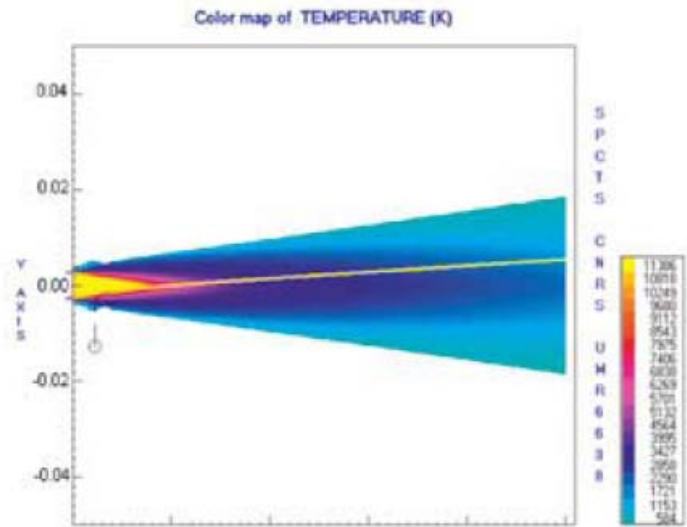

**Figure 5** - Thermal profile of plasma jet and particle path (yellow line).

**Table 4** - Properties of plasma gas in the bulk and on particle surface for the two studied temperatures.

| Gas temperature (K) | 10027 | 11357 |
|---|---|---|
| Particle surface temperature (K) | 1152 | 1766 |
| Reynolds number | 8.59 | 8.43 |
| Prandtl number | 0.29 | 0.43 |
| $\rho_\infty$ [kg/m³] | 0.0284 | 0.0250 |
| $\mu_\infty$ [Pa·s] | 2.42·10⁻⁴ | 2.37·10⁻⁴ |
| $\rho_{ps}$ [kg/m³] | 0.390 | 0.266 |
| $\mu_{ps}$ [Pa·s] | 5.65·10⁻⁵ | 7.59·10⁻⁵ |

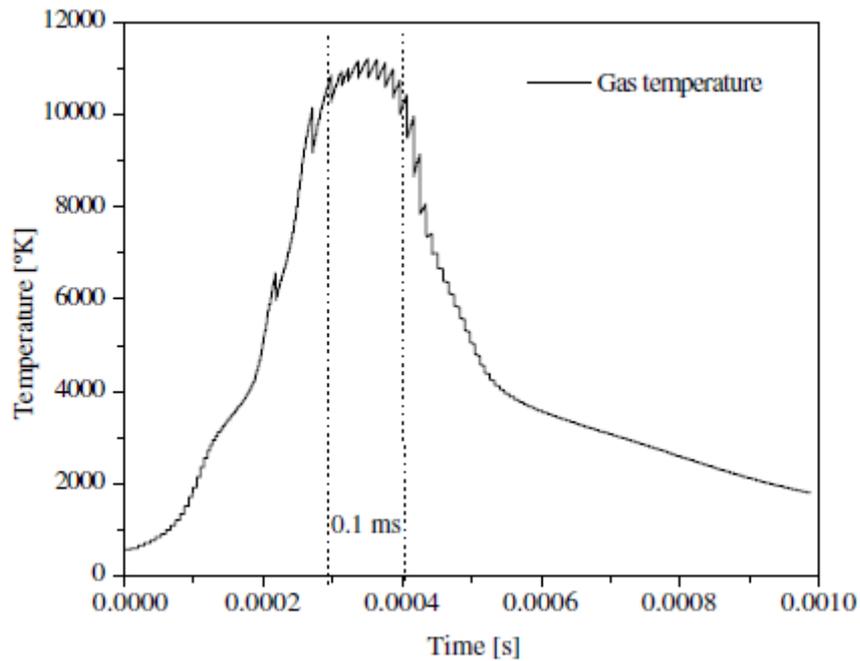

**Figure 6** - Thermal evolution of gas in the plasma jet.

**Figure 7** - (a) Mesh utilized in cases 1, 2 and 3; (b) Mesh utilized in case 4.

**Figure 8** - Initial conditions for cases 1 and 2.

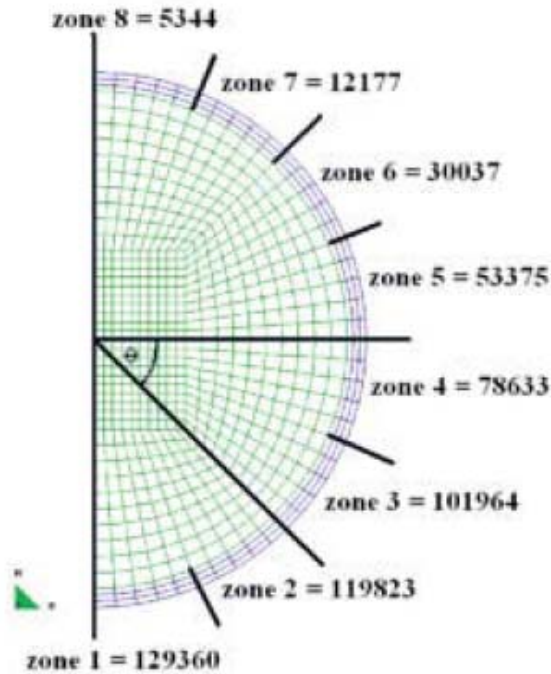

**Figure 9 - Distribution of the heat transfer coefficient along the particle surface for case 2.**

## RESULTS AND DISCUSSION

Figure 10 and Figure 11 show the temperature field computed for cases 1 and 2, respectively, at time 0.166 ms. From such temperature distributions the effect of uniform and non-uniform heat transfer coefficient is apparent. In Figure 10 (case 1) the thermal profile is constant along radial directions, meanwhile, the temperature varies along all the radios for case 2 (Figure 11). For both cases, the temperature of the covering layer is much higher than that of the center, due to the greater thermal resistance and melting point of the alumina. Simulations for these cases were carried out until the boiling point of alumina.

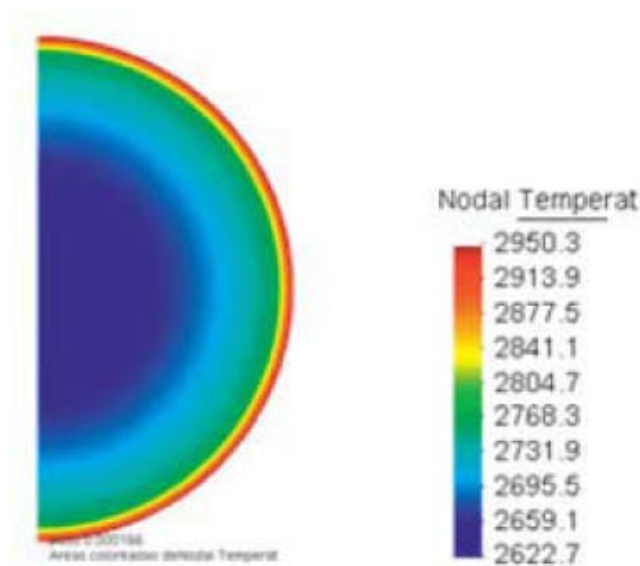

**Figure 10 - Thermal field of the particle obtained for case 1 at 0.166 ms.**

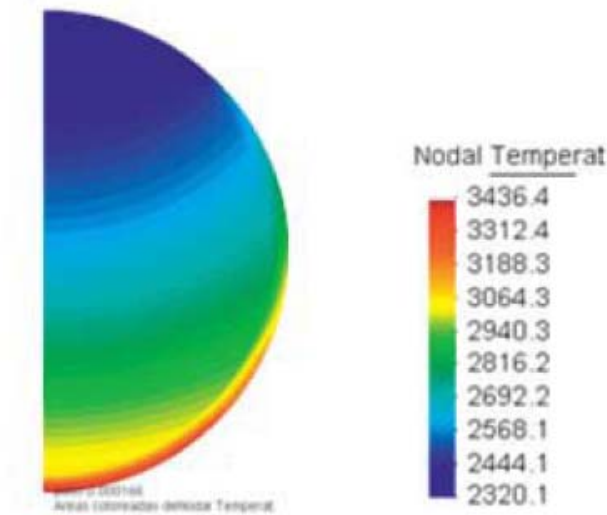

**Figure 11 - Thermal field of the particle obtained for case 2 at 0.166 ms.**

Figure 12 shows a comparison between the thermal evolution profiles obtained for cases 1 and 2 at the surface and center of the particle. The radial directions considered for points located on the particle surface in simulation 2 follow the $\hat{I}$, angles 0°, 45 ° and 315° with the horizontal axis, respectively; while for simulation 1, the thermal evolution profile is the same in the entire particle surface. It can be seen that most of the thermal evolutions increase at every time, except the point of the surface located at 315° which decreases until 0.005 ms and after increases following the same behavior than the others. This is due to its lower convection coefficient, which produces also a lower heat transfer from plasma gas to the particle. The phase change of the iron (center of particle) at 1810 K is well captured in the curves. The same phenomena for the alumina at 2326 K cannot be observed since the evolution is plotted at the surface where the heat transfer rapidly occurs due to the contact with the gas at very high temperature. In general, the temperature increases drastically inside the particle in a very short time (until 0.166 ms). Hence a complete melting of the metal (iron) and ceramic (alumina) rapidly occurs inside the particle.

Figure 13 shows a comparison between the thermal evolution profiles obtained for cases 1 and 2 at the surface and center of the particle, together with the results obtained in [1,31,32] using a thermal contact resistance "RTC" equals to $10^{-8}$ [m$^2$·K/W]. In general, the phase change time for iron (1810 K) obtained using our model is shorter than the obtained using the other formulation, however, numerical trends clearly agree with those obtained using a finite volume enthalpy based formulation.

**Figure 14** presents the temperature results obtained along a radial direction for case 3 using different constant values for the heat transfer coefficient. The thermal profiles increase for higher values of $h\infty$.

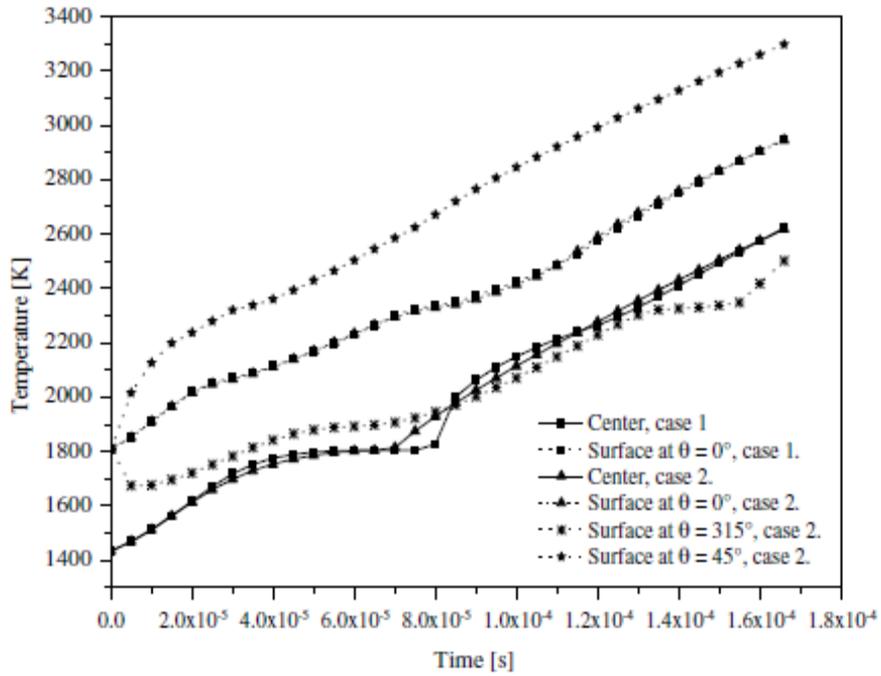

**Figure 12** Temperature evolution for cases 1 and 2 at points located at the surface and center of the particle.

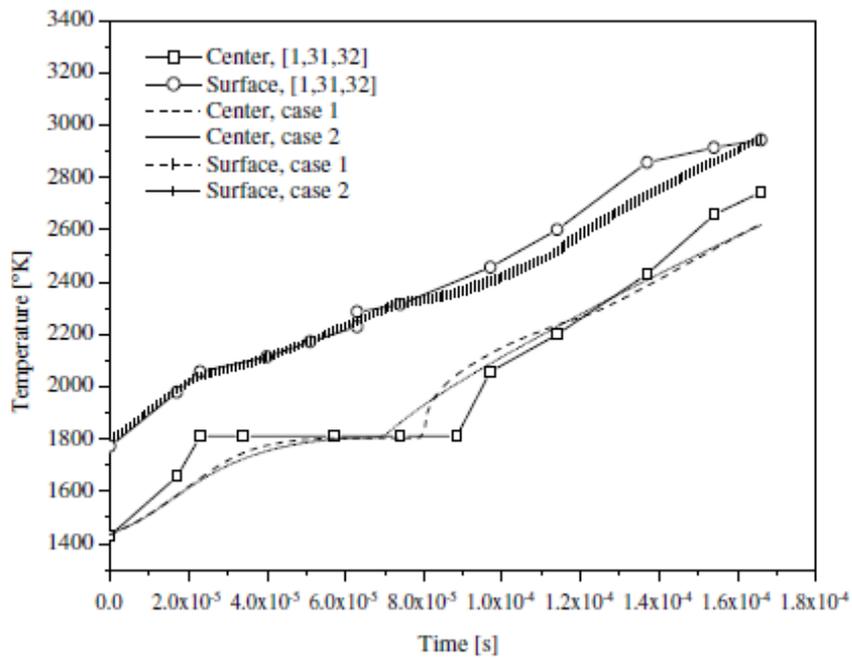

**Figure 13 -** Temperature evolution for cases 1 and 2 at points located at the surface and center of the particle in comparison with those obtained in [1,33,34]

Figure 15 and Figure 16 report the temperature field for cases 3 and 4 at 0.1 ms denoting the effect of the thickness of the alumina layer. In general, the thermal fields computed with a uniform heat transfer coefficient show similar trends to those obtained in [20].

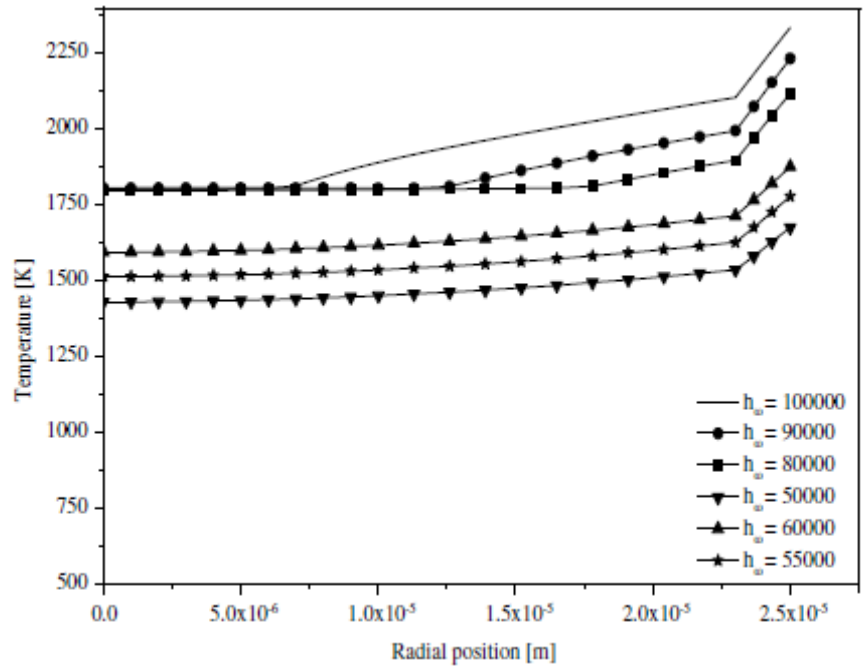

**Figure 14 -** Radial temperature profiles of the particle with a ratio $R_{Fe}/R_{particle} = 0.92$, for different values of $h_{âˆž}$ at 0.1ms in case 3.

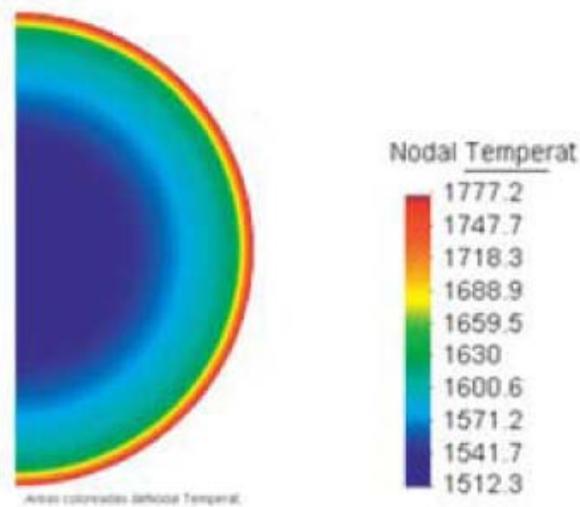

**Figure 15 -** Thermal field of the particle obtained for case 3 at 0.1 ms.

Figure 17 shows a comparison between thermal profiles along a radial direction for cases 3 and 4, together with results obtained in [1,31,32] using a finite volume enthalpy based formulation without thermal contact resistance "RTC". In general, the curves present two parts with different slopes corresponding to the different thermal conductivity coefficient of iron and ceramic. For the metal, the profile has a smaller slope due to the high thermal conductivity that always tends to eliminate the thermal gradient. In the ceramic part, the temperature profile exhibits a greater slope due its low thermal conductivity. This effect is more representative in the curve with a ratio $R_{Fe}/R_{particle}$ equals to 0.72, due to the higher ceramic layer thickness. Thus, it is very difficult to eliminate the temperature gradients generated inside a particle. In these both cases, the melting point of iron and alumina were not reached implying, as a consequence, that the heat transfer was only by conduction. The numerical trends obtained in this work clearly agree with those obtained using a finite volume enthalpy based formulation.

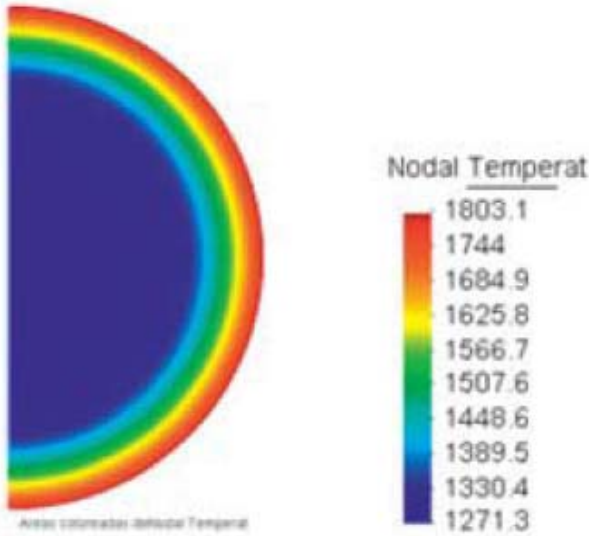

**Figure 16 - Thermal field of the particle obtained for case 4 at 0.1 ms.**

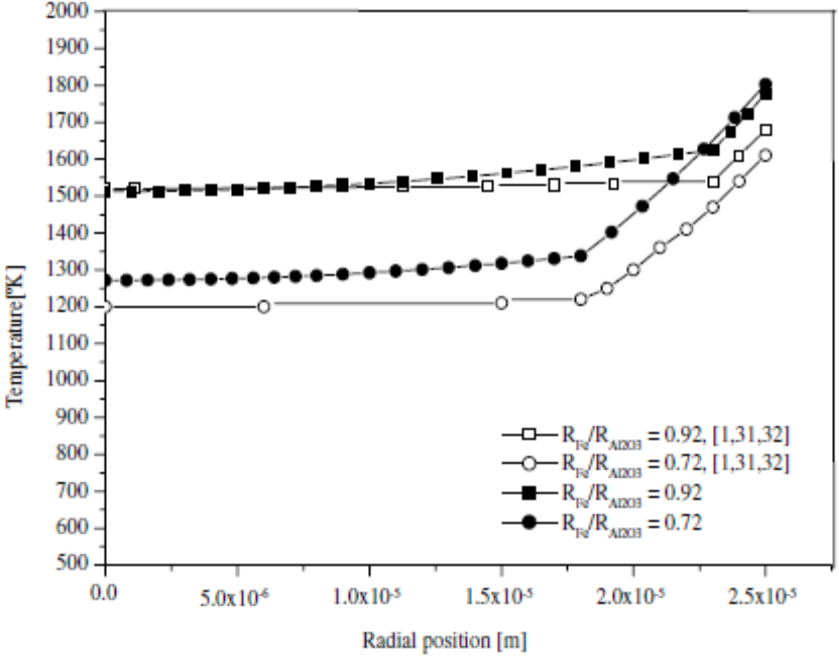

**Figure 17 - Radial thermal profiles of the particle obtained for cases 3 and 4 at 0.1ms in comparison with those obtained in [1,33,34].**

For all the cases simulated in this work, the temperature of the covering layer is higher than that of the center, due to the higher thermal resistance and melting point of the alumina.

## CONCLUSIONS

The problem of heat transfer between gas and particles inside a plasma jet was analyzed using a finite element temperature-based formulation.

The results obtained in the present work using this methodology satisfactorily described the melting process of the particle and thermal numerical trends clearly showed two parts with different slopes, corresponding to the different thermal conductivity coefficient of iron and ceramic in the bi-layer particle. Besides, these thermal numerical trends agreed with those previously reported in the literature and computed with a finite volume enthalpy based formulation.

## ACKNOLEDGMENTS

The authors thank the support provided by CNRS-CONICYT and project FONDECYT 1095028, in which the present work was developed.